\documentclass[reqno,amsmath]{amsart}
\usepackage[reqno]{amsmath}
\usepackage{graphicx}
\usepackage{fullpage}

\theoremstyle{definition}

\newcommand{\xmin}{x_{\rm min}}
\newcommand{\xmax}{x_{\rm max}}
\newcommand{\green}{f_{_{\rm G}}}
\newcommand{\sig}{\sigma_{_{\rm T}}}

\theoremstyle{remark}

\newcommand{\abs}[1]{\lvert#1\rvert}

\begin{document}

\title[Integrals of products of Whittaker functions]
{On the integration of products of Whittaker functions
with respect to the second index}

\author[Peter A. Becker]{Peter A. Becker\\
Center for Earth Observing and Space Research,\\
School of Computational Sciences,\\
George Mason University,\\
Fairfax, VA 22030-4444, USA}

\email{pbecker@gmu.edu}


\date{Submitted May 23, 2003.}


\vskip2.0truein
\centerline{Accepted for publication in the Journal of Mathematical
Physics}

\begin{abstract}
Several new formulas are developed that enable the evaluation of a
family of definite integrals containing the product of two Whittaker
$W_{\kappa,\,\mu}(x)$-functions. The integration is performed with
respect to the second index $\mu$, and the first index $\kappa$ is
permitted to have any complex value, within certain restrictions
required for convergence. The method utilizes complex contour
integration along with various symmetry relations satisfied by the
Whittaker functions. The new results derived in this paper are
complementary to the previously known integrals of products of Whittaker
functions, which generally treat integration with respect to either
the first index $\kappa$ or the primary argument $x$. A physical
application involving radiative transport is discussed.
\end{abstract}

\maketitle

\section*{\bf I. INTRODUCTION}
Many problems in mathematical physics involve differential equations
with solutions that can be expressed in terms of Whittaker's functions
$W_{\kappa,\,\mu}(x)$ and $M_{\kappa,\,\mu}(x)$. Examples of the diverse
applications include studies of the spectral evolution resulting from
the Compton scattering of radiation by hot electrons,$^{1,2,3}$ modeling
of the structure of the hydrogen atom,$^{4}$ analysis of the
Schr\"odinger equation,$^{5}$ studies of the Coulomb Green's
function,$^{6}$ and analysis of fluctuations in financial markets.$^{7}$

In a number of applications, it is necessary to evaluate integrals of
Whittaker functions. This need may arise out of the requirement to
satisfy normalization or orthogonality conditions. In particular, in the
analysis of time-dependent Compton scattering, it is necessary to
evaluate integrals containing the product of two Whittaker
$W_{\kappa,\,\mu}(x)$-functions, where the variable of integration is
the second index $\mu$. This is an unusual situation that is not covered
by any of the previously known formulas for integrals of products of
Whittaker functions. The required integrals in the Compton scattering
application are members of the general family
\begin{equation}
I(s) \equiv \int_0^\infty
{u \, \sinh(2 \pi u) \, \Gamma(1/2-\kappa-i u)
\, \Gamma(1/2-\kappa+i u) \over s + u^2}
\, W_{\kappa , \, i u}(x)
\, W_{\kappa , \, i u}(x_0) \, du \ ,
\label{eq1}
\end{equation}
where $x$ and $x_0$ are real and positive, and $s$ and $\kappa$ are
complex. This integral converges for all values of $s$ in the complex
plane, with the exclusion of the negative real semiaxis, provided that
${\mathcal Re} \, \kappa \ne {1 \over 2}, {3 \over 2}, {5 \over 2},
\ldots$, if ${\mathcal Im} \, \kappa \ne 0$. It also converges in the
special case $s=0$, provided ${\mathcal Re} \, \kappa \neq {1 \over 2},
{3 \over 2}, {5 \over 2}, \ldots$ In this paper we derive several exact
formulas for the evaluation of the integral $I(s)$ that fully describe
all of the convergent cases.

\section*{\bf II. FUNDAMENTAL EQUATIONS}

We shall begin by briefly reviewing some of the basic properties of the
Whittaker functions that will be useful in our later work. The Whittaker
functions $W_{\kappa,\,\mu}(z)$ and $M_{\kappa,\,\mu}(z)$ are confluent
hypergeometric functions that are related to the Kummer functions
$\Phi(a,b,z)$ and $\Psi(a,b,z)$ by$^{8,9}$
\begin{equation}
\begin{split}
M_{\kappa,\,\mu}(z) = z^{\mu+1/2} \, e^{-z/2} \, \Phi(1/2+\mu-\kappa,
\, 1+2\mu; \, z) \ , \\
W_{\kappa,\,\mu}(z) = z^{\mu+1/2} \, e^{-z/2} \, \Psi(1/2+\mu-\kappa,
\, 1+2\mu; \, z) \ .
\end{split}
\label{eq2}
\end{equation}
For small values of $\abs{z}$, the function $M_{\kappa, \, \mu}(z)$ is
given by the power series
\begin{equation}
M_{\kappa,\,\mu}(z) = e^{-z/2} \, z^{\mu + 1/2} \, \sum_{n=0}^\infty
{(1/2-\kappa+\mu)_n \over (1 + 2\mu)_n} \, {z^n \over n!} \ ,
\label{eq3}
\end{equation}
where $(a)_n$ denotes the Pochhammer symbol, defined by$^9$
\begin{equation}
(a)_n \equiv {\Gamma(a+n) \over \Gamma(a)} \ .
\label{eq4}
\end{equation}
The function $W_{\kappa,\,\mu}(z)$ can be expressed in terms of
$M_{\kappa,\,\mu}(z)$ using$^8$
\begin{equation}
W_{\kappa,\,\mu}(z) = {\Gamma(-2\mu) \over \Gamma(1/2-\mu-\kappa)}
\, M_{\kappa,\,\mu}(z)
+ {\Gamma(2\mu) \over \Gamma(1/2+\mu-\kappa)}
\, M_{\kappa,\,-\mu}(z) \ .
\label{eq5}
\end{equation}

The integrand in equation (\ref{eq1}) for $I(s)$ is an even function
of $u$, and therefore we can write
\begin{equation}
I(s) = {1 \over 2} \int_{-\infty}^\infty
{u \, \sinh(2 \pi u) \, \Gamma(1/2-\kappa-i u)
\, \Gamma(1/2-\kappa+i u) \over s + u^2}
\, W_{\kappa , \, i u}(x)
\, W_{\kappa , \, i u}(x_0) \, du \ .
\label{eq6}
\end{equation}
Next we utilize (\ref{eq5}) to express $W_{\kappa,\,iu}(x_0)$ as
\begin{equation}
W_{\kappa,\,iu}(x_0) = {\Gamma(-2\,iu) \over \Gamma(1/2-\kappa-iu)}
\, M_{\kappa,\,iu}(x_0)
+ {\Gamma(2\,iu) \over \Gamma(1/2-\kappa+iu)}
\, M_{\kappa,\,-iu}(x_0) \ ,
\label{eq7}
\end{equation}
which can be rewritten as
\begin{multline}
\hskip0.5truein
W_{\kappa,\,iu}(x_0) = {\Gamma(-2\,iu) \, \Gamma(2\,iu) \over
\Gamma(1/2-\kappa-iu) \, \Gamma(1/2-\kappa+iu)} \\
\times \ \bigg[
{\Gamma(1/2-\kappa+iu) \over \Gamma(2\,iu)} \, M_{\kappa,\,iu}(x_0)
+ {\Gamma(1/2-\kappa-iu) \over \Gamma(-2\,iu)} \, M_{\kappa,\,-iu}(x_0)
\bigg] \ .
\hskip0.5truein
\label{eq8}
\end{multline}
By employing the recurrence formula for the gamma function, $z \, \Gamma(z)
=\Gamma(z+1)$, we can obtain the alternative form
\begin{multline}
\hskip0.5truein
W_{\kappa,\,iu}(x_0) = {\Gamma(-2\,iu) \, \Gamma(1+2\,iu)
\over \Gamma(1/2-\kappa-iu) \, \Gamma(1/2-\kappa+iu)} \\
\times \ \bigg[
{\Gamma(1/2-\kappa+iu) \over \Gamma(1+2\,iu)} \, M_{\kappa,\,iu}(x_0)
- {\Gamma(1/2-\kappa-iu) \over \Gamma(1-2\,iu)} \, M_{\kappa,\,-iu}(x_0)
\bigg] \ .
\hskip0.5truein
\label{eq9}
\end{multline}
Using this result to substitute for $W_{\kappa,\,iu}(x_0)$ in
(\ref{eq6}) now yields
\begin{multline}
\hskip0.5truein
I(s) = {1 \over 2} \int_{-\infty}^\infty {u \, \sinh(2 \pi u) \over s+u^2}
\, \Gamma(-2\,iu) \, \Gamma(1+2\,iu) \, W_{\kappa,\,iu}(x) \\
\times
\ \left[
{\Gamma(1/2-\kappa+iu) \over \Gamma(1+2\,iu)} \, M_{\kappa,\,iu}(x_0)
- {\Gamma(1/2-\kappa-iu) \over \Gamma(1-2\,iu)} \, M_{\kappa,\,-iu}(x_0)
\right] du \ .
\hskip0.5truein
\label{eq10}
\end{multline}
By utilizing the reflection formula for the gamma function,
\begin{equation}
\Gamma(1+2\,iu) \, \Gamma(-2\,iu) = {\pi \, i \over \sinh(2 \pi u)}
\ ,
\label{eq11}
\end{equation}
along with the symmetry relation [see Eq.~(\ref{eq5})]
\begin{equation}
W_{\kappa,\,iu}(x) = W_{\kappa,\,-iu}(x) \ ,
\label{eq12}
\end{equation}
we can rewrite (\ref{eq10}) as
\begin{multline}
\hskip0.5truein
I(s) = {\pi i \over 2} \int_{-\infty}^\infty
{u \over s+u^2} \bigg[
{\Gamma(1/2-\kappa+iu) \over \Gamma(1+2\,iu)} \,
W_{\kappa,\,iu}(x) \, M_{\kappa,\,iu}(x_0) \\
- {\Gamma(1/2-\kappa-iu) \over \Gamma(1-2\,iu)} \,
W_{\kappa,\,-iu}(x) \, M_{\kappa,\,-iu}(x_0)
\bigg] du \ .
\hskip0.5truein
\label{eq13}
\end{multline}
This relation can be split into two identical integrals, and
consequently our expression for $I(s)$ can be reduced to
\begin{equation}
\hskip0.5truein
I(s) = - \pi i \int_{-\infty}^\infty
{u \over s+u^2} {\Gamma(1/2-\kappa-iu) \over \Gamma(1-2\,iu)} \,
W_{\kappa,\,-iu}(x) \, M_{\kappa,\,-iu}(x_0) \, du \ .
\label{eq14}
\end{equation}

\section*{\bf III. CONTOUR INTEGRATION}

The fundamental expression for the integral $I(s)$ given by (\ref{eq1})
is clearly symmetrical with respect to the interchange of $x$ and $x_0$.
We can use this flexibility to select the arguments of the $W$ and $M$
functions in such a way that the integration along the curved portion of
the closed contour $C$ in Fig.~1 vanishes in the limit $r \to \infty$.
By employing asymptotic analysis, we find that this occurs if $\xmax$ is
the argument of the $W$ function and $\xmin$ is the argument of the $M$
function, where
\begin{equation}
\xmin \equiv \min(x,x_0) \ , \ \ \ \ \ \ 
\xmax \equiv \max(x,x_0) \ .
\label{eq15}
\end{equation}
Equation~(\ref{eq14}) for $I(s)$ can therefore be recast as the
complex contour integral
\begin{equation}
I(s) = \oint_{C} L(u) \, du \ ,
\label{eq16}
\end{equation}
where
\begin{equation}
L(u) \equiv - \pi \, i \, {u \over s+u^2} \,
{\Gamma(1/2-\kappa-iu) \over
\Gamma(1-2\,iu)} \ W_{\kappa,\,-iu}(\xmax) \, M_{\kappa,\,-iu}(\xmin)
\ .
\label{eq17}
\end{equation}
We shall proceed to obtain an exact, closed form expression for
$I(s)$ by utilizing the residue theorem to evaluate the integral
in (\ref{eq16}).

\begin{figure}
\hspace{5mm}
\includegraphics[width=100mm]{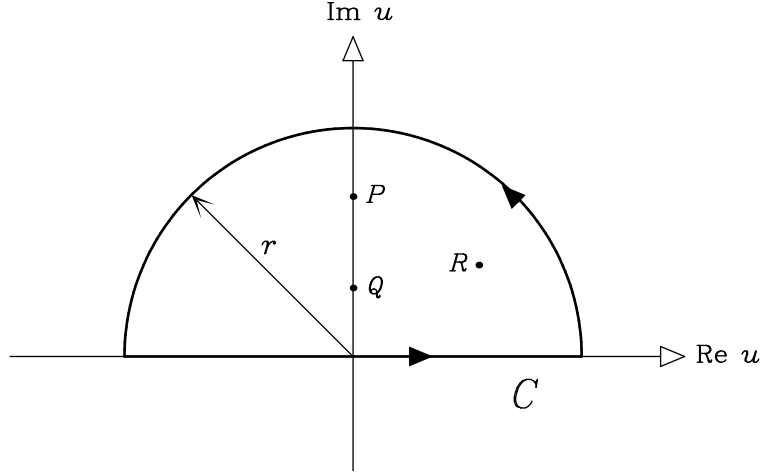}
\caption{Integration around the closed contour $C$ yields $I(s)$
in the limit $r \to \infty$ [see Eq.~(\ref{eq16})]. In this example,
$\kappa$ is a real number in the range $3/2 < \kappa < 5/2$, and consequently
there are two simple poles ($P$ and $Q$) located inside the contour on
the imaginary axis [see Eq.~(\ref{eq18})]. The imaginary part of $s$ is
less than zero in this instance, and consequently there is also a simple
pole, $R$, located at $u=i\sqrt{s}$ in quadrant~I.}
\end{figure}

The integrand $L(u)$ has a simple pole located at $u=i\sqrt{s}$, where
$\sqrt{s}$ denotes the principle branch of the square root function.
This pole is located in quadrant~II of the complex $u$ plane if ${\mathcal
Im} \, s \ge 0$, and otherwise it is located in quadrant~I. In either
case, the pole is contained within the closed integration contour $C$.
Additional simple poles are located at the singularities of the function
$\Gamma(1/2-\kappa-iu)$, which occur where the quantity $1/2-\kappa-iu\,$
is equal to zero or a negative integer. At least one of the poles falls
in the upper half-plane if ${\mathcal Re} \, \kappa > 1/2$. The poles
are located at $u=u_n$, where
\begin{equation}
u_n \equiv \, i \left(\kappa - {1 \over 2} - n\right) \ ,
\ \ \ \ \ n = 0, 1, \ldots, \lbrack {\mathcal Re} \,
\kappa - {1 \over 2}\rbrack \ ,
\label{eq18}
\end{equation}
and $\lbrack a\rbrack$ indicates the integer part of $a$. Note that
if ${\mathcal Re} \, \kappa < 1/2$, then only the pole at $u=i\sqrt{s}$
is contained within the contour $C$.

We can now use the residue theorem to write
\begin{equation}
I(s) = \ 2 \pi i \sum_{n=0}^{\lbrack {\mathcal Re} \,
\kappa - {1 \over 2} \rbrack} \, {\rm Res}(u_n)
\, + \, 2 \pi i \ {\rm Res}(i\sqrt{s}) \ ,
\label{eq19}
\end{equation}
where ${\rm Res}(u_*)$ denotes the residue associated with the simple
pole located at $u = u_*$.

\section*{\bf IV. EVALUATION OF THE RESIDUES}

The residue corresponding to the simple pole at $u=i\sqrt{s}$ is easily
computed using the formula
\begin{equation}
{\rm Res}(i\sqrt{s}) = \lim_{u \to i\sqrt{s}} \ (u - i\sqrt{s})
\, L(u) \ ,
\label{eq20}
\end{equation}
which can be immediately evaluated to obtain
\begin{equation}
{\rm Res}(i\sqrt{s}) = - {\pi i \over 2} \, {\Gamma(1/2-\kappa+\sqrt{s})
\over \Gamma(1+2\sqrt{s})} \ W_{\kappa,\sqrt{s}}\,(\xmax) \,
M_{\kappa,\sqrt{s}}\,(\xmin)
\ .
\label{eq21}
\end{equation}
Similarly, the residues associated with the simple poles located at
$u=u_n$ are evaluated using
\begin{equation}
{\rm Res}(u_n) = \lim_{u \to u_n} \ (u - u_n) \, L(u) \ .
\label{eq22}
\end{equation}
Because the poles in this case correspond to the singularities of the
function $\Gamma(1/2-\kappa-iu)$, we will require evaluation of the
quantity
\begin{equation}
\lim_{u \to u_n} \ (u - u_n) \ \Gamma\left({1 \over 2} - \kappa
- i u \right) \ .
\label{eq23}
\end{equation}
By combining (\ref{eq4}) and (\ref{eq18}) with the recurrence relation
$z \, \Gamma(z) = \Gamma(z+1)$, we obtain
\begin{equation}
\Gamma\left({1 \over 2} - \kappa - i u \right)
= {\Gamma(1/2 - \kappa - i u + n) \over
(1/2 - \kappa - i u)_n}
= {i \, \Gamma(1 + i u_n - i u) \over
(i u_n - i u - n)_n \, (u - u_n)} \ ,
\label{eq24}
\end{equation}
and therefore
\begin{equation}
\lim_{u \to u_n} \ (u - u_n) \ \Gamma\left({1 \over 2} - \kappa
- i u \right)
= {\, i \, (-1)^n \over n!} \ ,
\label{eq25}
\end{equation}
where we have used the fact that $(-n)_n = (-1)^n n!$ Next we need to
evaluate the Whittaker functions appearing on the right-hand side of
(\ref{eq17}) in the limit $u \to u_n$. Using (\ref{eq2}) and
(\ref{eq18}), we find that
\begin{equation}
\begin{split}
M_{\kappa,\,-iu_n}\,(z) =  e^{-z/2} \, z^{\kappa-n}
\ \Phi(-n, \, 2\kappa-2n; \, z) \ , \\
W_{\kappa,\,-iu_n}\,(z) = e^{-z/2} \, z^{\kappa-n}
\ \Psi(-n, \, 2\kappa-2n; \, z) \ .
\end{split}
\label{eq26}
\end{equation}
By employing equations~(13.6.9) and (13.6.27) from Abramowitz and
Stegun,$^9$ we can rewrite these expressions as
\begin{equation}
\begin{split}
M_{\kappa,\,-iu_n}\,(z) =  {n! \over (\alpha+1)_n} \, e^{-z/2} \,
z^{(\alpha+1)/2} \, P^{(\alpha)}_n(z) \ , \\
W_{\kappa,\,-iu_n}\,(z) = (-1)^n \, n! \ e^{-z/2} \,
z^{(\alpha+1)/2} \, P^{(\alpha)}_n(z) \ ,
\end{split}
\label{eq27}
\end{equation}
where $P^{(\alpha)}_n(z)$ denotes the Laguerre polynomial, and
\begin{equation}
\alpha \equiv 2 \kappa - 2 n - 1 = - 2 \, i \, u_n \ .
\label{eq28}
\end{equation}
Combining (\ref{eq17}), (\ref{eq22}), (\ref{eq25}), and (\ref{eq27}),
we obtain for the residue
\begin{equation}
{\rm Res}(u_n) = {2 \, \pi \, i \, \alpha \over 4 s - \alpha^2} \,
{n! \over \Gamma(\alpha+n+1)}
\ e^{-(x+x_0)/2} \, (x \, x_0)^{(\alpha+1)/2}
\, P^{(\alpha)}_n(x) \ P^{(\alpha)}_n(x_0) \ .
\label{eq29}
\end{equation}
Utilizing this result along with (\ref{eq19}) and (\ref{eq21}),
we conclude that
\begin{equation}
\begin{split}
I(s) = \int_0^\infty
{u \, \sinh(2 \pi u) \, \Gamma(1/2-\kappa-i u)
\, \Gamma(1/2-\kappa+i u) \over s + u^2}
\, W_{\kappa , \, i u}(x)
\, W_{\kappa , \, i u}(x_0) \, du \hskip1.0truein \\
= \pi^2 \, {\Gamma(1/2-\kappa+\sqrt{s}) \over
\Gamma(1 + 2 \sqrt{s})} \ W_{\kappa,\sqrt{s}}\,(\xmax)
\ M_{\kappa,\sqrt{s}}\,(\xmin) \hskip1.5truein \\
- \, 4 \pi^2 \, e^{-(x+x_0)/2} \, \sum_{n=0}^{\lbrack
{\mathcal Re} \, \kappa - {1 \over 2} \rbrack}
\, {\alpha \, n! \over \Gamma(\alpha+n+1)}
\, {(x \, x_0)^{(\alpha+1)/2} \over 4 s - \alpha^2}
\, P^{(\alpha)}_n(x) \ P^{(\alpha)}_n(x_0) \ ,
\end{split}
\label{eq30}
\end{equation}
where $\alpha = 2 \kappa - 2 n - 1$. This previously unknown integral
formula is one of the main results of the paper. Note that the summation
is carried out only if ${\mathcal Re} \, \kappa \geq 1/2$. The integral
on the left-hand side of (\ref{eq30}) converges for all complex values
of $s$ with the exception of the negative real semiaxis, provided that
${\mathcal Re} \, \kappa \ne {1 \over 2}, {3 \over 2}, {5 \over 2},
\ldots$, if ${\mathcal Im} \, \kappa \ne 0$. When $s=0$, the integral
converges provided ${\mathcal Re} \, \kappa$ is not a positive half-integer.

A case of special interest can be generated by setting $x_0 = x$ in
(\ref{eq30}). The result obtained is the {\it quadratic normalization
integral},
\begin{equation}
\begin{split}
\int_0^\infty
{u \, \sinh(2 \pi u) \, \Gamma(1/2-\kappa-i u)
\, \Gamma(1/2-\kappa+i u) \over s + u^2}
\, W^2_{\kappa , \, i u}(x) \, du \hskip1.0truein \\
= \pi^2 \, {\Gamma(1/2-\kappa+\sqrt{s}) \over
\Gamma(1 + 2 \sqrt{s})} \ W_{\kappa,\sqrt{s}}\,(x)
\ M_{\kappa,\sqrt{s}}\,(x) \hskip1.5truein \\
- \, 4 \pi^2 \, e^{-x} \, \sum_{n=0}^{\lbrack
{\mathcal Re} \, \kappa - {1 \over 2} \rbrack}
\, {\alpha \, n! \over \Gamma(\alpha+n+1)}
\, {x^{\alpha+1} \over 4 s - \alpha^2}
\, \left[P^{(\alpha)}_n(x)\right]^2 \ ,
\end{split}
\label{eq31}
\end{equation}
which is useful in situations involving the development of a series
expansion in terms of a set of normalized basis functions. In the
following sections, we shall proceed to discuss the limiting behavior of
(\ref{eq30}) observed when two of the poles coincide, as well as its
relation to formulas appearing in the previous literature.

\section*{\bf V. LIMITING BEHAVIOR}

An interesting situation arises if the quantity $1/2-\kappa+\sqrt{s}\,$
is equal to zero or a negative integer, because in this case the integral
$I(s)$ converges, although the first term on the right-hand side of
(\ref{eq30}) formally {\it diverges} due to the appearance of the
factor $\Gamma(1/2-\kappa+\sqrt{s})$. This occurs when
\begin{equation}
\sqrt{s} = \sqrt{s_m} \equiv \kappa - {1 \over 2} - m \ ,
\label{eq32}
\end{equation}
where $m$ is a positive integer or zero. Since $\sqrt{s}$ denotes the
principle branch of the square root function, it follows that $\sqrt{s}$
is located in either quadrants~I or IV of the complex $s$ plane,
depending on whether ${\mathcal Im} \, s$ is positive or negative. Hence
${\mathcal Re} \, \sqrt{s} \ge 0$ in general, and therefore the function
$\Gamma(1/2-\kappa+\sqrt{s})$ has no singularities unless ${\mathcal Re}
\, \kappa \ge 1/2$. The values of $m$ yielding singularities for a given
value of $\kappa$ are
\begin{equation}
m = 0, 1, \ldots, \lbrack {\mathcal Re} \, \kappa - {1 \over 2}
\rbrack \ .
\label{eq33}
\end{equation}

When $s = s_m$, the divergence of the first term on the right-hand side
of (\ref{eq30}), containing the factor $\Gamma(1/2-\kappa+\sqrt{s})$, is
exactly balanced by the divergence of the $n = m$ term in the sum,
leaving a finite residual quantity. This situation corresponds to a
coincidence of the pole located at $u=i\sqrt{s}$ with the pole located
at $u=u_m=i(\kappa-1/2-m)$ [see Eq.~(\ref{eq18})]. In this case the
resulting pole has order two. The associated residue can be computed by
using the standard formula for a second-order pole, but it is more
efficient to approach the calculation by evaluating equation~(\ref{eq30})
for $I(s)$ in the limit $s \to s_m$. The limiting value of the sum of the
two divergent terms is given by
\begin{equation}
\begin{split}
K \equiv \lim_{s \to s_m}
\ \pi^2 \, {\Gamma(1/2-\kappa+\sqrt{s}) \over
\Gamma(1 + 2 \sqrt{s})} \ W_{\kappa,\sqrt{s}}\,(\xmax)
\ M_{\kappa,\sqrt{s}}\,(\xmin) \hskip1.5truein \\
- \, 4 \pi^2 \, e^{-(x+x_0)/2} \, {\lambda \, m! \over
\Gamma(\lambda + m + 1)} \, {(x \, x_0)^{(\lambda+1)/2}
\over 4 s - \lambda^2} \, P^{(\lambda)}_m(x) \ P^{(\lambda)}_m(x_0)
\ ,
\end{split}
\label{eq34}
\end{equation}
where
\begin{equation}
\lambda \equiv 2 \kappa - 2 m - 1 = 2 \, \sqrt{s_m} \ .
\label{eq35}
\end{equation}
Equation~(\ref{eq34}) can be rewritten as
\begin{equation}
K = \lim_{s \to s_m} {N \over D} \ ,
\label{eq36}
\end{equation}
where
\begin{equation}
\begin{split}
N \equiv
\pi^2 \, (s - s_m\,){\Gamma(1/2-\kappa+\sqrt{s}) \over
\Gamma(1 + 2 \sqrt{s})} \ W_{\kappa,\sqrt{s}}\,(\xmax)
\ M_{\kappa,\sqrt{s}}\,(\xmin) \hskip0.5truein \\
- \, \pi^2 \, e^{-(x+x_0)/2} \, {\lambda \, m! \over
\Gamma(\lambda + m + 1)} \, (x \, x_0)^{(\lambda+1)/2}
\, P^{(\lambda)}_m(x) \ P^{(\lambda)}_m(x_0)
\ ,
\end{split}
\label{eq37}
\end{equation}
and
\begin{equation}
D \equiv s - s_m \ .
\label{eq38}
\end{equation}

We can demonstrate that the numerator $N$ vanishes in the limit
$s \to s_m$ as follows. First we use (\ref{eq4}) and (\ref{eq32})
along with the recurrence relation for the gamma function to write
\begin{equation}
\Gamma\left({1 \over 2} - \kappa + \sqrt{s} \right)
= {\Gamma(1/2 - \kappa + \sqrt{s} + m) \over
(1/2 - \kappa + \sqrt{s})_m}
= {(\sqrt{s} + \sqrt{s_m}) \ \Gamma(1 + \sqrt{s} - \sqrt{s_m})
\over (s - s_m) \, (\sqrt{s} - \sqrt{s_m} - m)_m} \ ,
\label{eq39a}
\end{equation}
and therefore [cf. Eq.~(\ref{eq25})]
\begin{equation}
\lim_{s \to s_m} \ (s - s_m) \ \Gamma\left({1 \over 2} - \kappa
+ \sqrt{s}\right)
= {(-1)^m \over m!} \, 2 \sqrt{s_m} \ .
\label{eq39}
\end{equation}
Furthermore, based on (\ref{eq27}), (\ref{eq28}), and (\ref{eq35}),
we note that
\begin{equation}
\begin{split}
M_{\kappa,\sqrt{s_m}}\,(z) =  {m! \over (\lambda+1)_m} \, e^{-z/2} \,
z^{(\lambda+1)/2} \, P^{(\lambda)}_m(z) \ , \\
W_{\kappa,\sqrt{s_m}}\,(z) = m! \, (-1)^m \, e^{-z/2} \,
z^{(\lambda+1)/2} \, P^{(\lambda)}_m(z) \ .
\end{split}
\label{eq40}
\end{equation}
Taken together, (\ref{eq37}), (\ref{eq39}), and (\ref{eq40}) indicate
that the numerator $N$ vanishes in the limit $s \to s_m$. The
denominator $D$ also vanishes in this limit, and therefore we can employ
L'H\^opital's rule to evaluate $K$ by writing
\begin{equation}
K = \lim_{s \to s_m} \, {\partial N \over \partial s}
\ \bigg/
\lim_{s \to s_m} \, {\partial D \over \partial s}
\ .
\label{eq41}
\end{equation}
Since $\partial D/\partial s = 1$ and the second term on the right-hand
side of (\ref{eq37}) is independent of $s$, we obtain
\begin{equation}
K = \lim_{s \to s_m} \, {\partial \over \partial s}
\ \pi^2 \, (s - s_m\,){\Gamma(1/2-\kappa+\sqrt{s}) \over
\Gamma(1 + 2 \sqrt{s})} \ W_{\kappa,\sqrt{s}}\,(\xmax)
\ M_{\kappa,\sqrt{s}}\,(\xmin)
\ .
\label{eq42}
\end{equation}
Upon differentiation, we obtain after a fairly lengthly calculation
\begin{equation}
K = {\pi^2 \, e^{-(x+x_0)/2} \, (x \, x_0)^{(\lambda+1)/2} \, m!
\, P^{(\lambda)}_m(x) \ P^{(\lambda)}_m(x_0) \over \Gamma(\lambda+m+1)}
\, \left[-\gamma_{_{\rm E}} + {1 \over \lambda} + H
- 2 \, \psi(\lambda+1)- {1 \over m+1} + \sum_{n=1}^{m+1} {1 \over n}
\right]
\ ,
\label{eq43}
\end{equation}
where $\gamma_{_{\rm E}} \approx -0.577$ is Euler's constant,
$\lambda = 2 \kappa - 2 m - 1$,
\smallskip
\begin{equation}
H \equiv
\, {\partial \over \partial\beta} \ln\left[W_{\kappa,\,\beta}\,(\xmax) \,
M_{\kappa,\,\beta}\,(\xmin) \right]\bigg|_{\beta=\sqrt{s_m}}
\ ,
\label{eq44}
\end{equation}
and
\begin{equation}
\psi(z) \equiv {d \over dz} \, \ln \Gamma(z) \ .
\label{eq45}
\end{equation}
Combining results, we find that in the special case $s = s_m
= (\kappa - m - 1/2)^2$ the integral $I(s)$ is given by
\begin{equation}
\begin{split}
I(s) = \int_0^\infty
{u \, \sinh(2 \pi u) \, \Gamma(1/2-\kappa-i u)
\, \Gamma(1/2-\kappa+i u) \over (\kappa-m-1/2)^2 + u^2}
\, W_{\kappa , \, i u}(x)
\, W_{\kappa , \, i u}(x_0) \, du \hskip1.0truein \\
= K - \, 4 \pi^2 \, e^{-(x+x_0)/2} \, \sum_{n=0 \atop n \neq m}^{\lbrack
{\mathcal Re} \, \kappa - {1 \over 2} \rbrack}
\, {\alpha \, n! \over \Gamma(\alpha+n+1)}
\, {(x \, x_0)^{(\alpha+1)/2} \over 4 s - \alpha^2}
\, P^{(\alpha)}_n(x) \ P^{(\alpha)}_n(x_0) \ ,
\end{split}
\label{eq46}
\end{equation}
where $\alpha = 2 \kappa - 2 n - 1$. The allowed range of values for $m$
is given by (\ref{eq33}), which indicates that we must have ${\mathcal Re}
\, \kappa \ge 1/2$ in order for any of these special cases to occur. Note
that the singular term with $n=m$ is not included in the sum, since that
term is contained within $K$. Equations~(\ref{eq30}) and (\ref{eq46})
cover all of the convergent cases of the fundamental integral $I(s)$. In
Sec.~VI we present simplified results obtained for certain values of the
parameters.

\section*{\bf VI. SPECIAL CASES}

The general nature of the expression for $I(s)$ given by (\ref{eq30})
encompasses many interesting special cases involving particular values
for the parameters $\kappa$, $s$, $x$, and $x_0$. In this section, we
shall briefly discuss a few illustrative examples obtained when the
first index $\kappa$ is equal to an integer, in which case the general
solution for $I(s)$ simplifies considerably. For brevity, we shall focus
here on situations with $s \neq s_m$. However, we emphasize that
formulas similar to those discussed below that are applicable to the
case $s = s_m$ can also be obtained in a straightforward manner by
starting with (\ref{eq46}) rather than (\ref{eq30}).

\vskip0.2truein
\centerline{\bf A. $\boldsymbol{\kappa = 0}$}
\vskip0.2truein

When $\kappa = 0$, the summation in (\ref{eq30}) is not performed at all.
Making use of the identities$^9$
\begin{equation}
\Gamma\left({1 \over 2} - i u \right) \,
\Gamma\left({1 \over 2} + i u \right) = {\pi \over \cosh(\pi u)}
\ ,
\label{eq47}
\end{equation}
and
\begin{equation}
\sinh(2 \pi u) = 2 \sinh(\pi u) \, \cosh(\pi u)
\ ,
\label{eq48}
\end{equation}
we find that (\ref{eq30}) reduces to
\begin{equation}
\int_0^\infty
{u \, \sinh(\pi u) \over s + u^2}
\, W_{0 , \, i u}(x)
\, W_{0 , \, i u}(x_0) \, du
= {\pi \over 2} \, {\Gamma(1/2+\sqrt{s}) \over
\Gamma(1 + 2 \sqrt{s})} \ W_{0,\sqrt{s}}\,(\xmax)
\ M_{0,\sqrt{s}}\,(\xmin) \ .
\label{eq49}
\end{equation}
This result is convergent for all complex values of $s$, excluding
the negative real semiaxis. Hence the point $s=0$ is convergent
in this case.

\vskip0.2truein
\centerline{\bf B. $\boldsymbol{\kappa = 1}$}
\vskip0.2truein

When $\kappa=1$, there is one simple pole located at $s_0 = 1/4$, and
we can make use of the identity
\begin{equation}
\Gamma\left(-{1 \over 2} - i u \right) \,
\Gamma\left(-{1 \over 2} + i u \right) = {4 \pi \over \cosh(\pi u)
\, (1 + 4 u^2)}
\ ,
\label{eq50}
\end{equation}
along with (\ref{eq48}) to reduce (\ref{eq30}) to the form
\begin{equation}
\begin{split}
\int_0^\infty
{u \, \sinh(\pi u) \over (1 + 4 u^2) \, (s + u^2)}
\, W_{1 , \, i u}(x)
\, W_{1 , \, i u}(x_0) \, du \hskip2.0truein \\
= {\pi \over 8} \, {\Gamma(\sqrt{s}-1/2) \over
\Gamma(1 + 2 \sqrt{s})} \ W_{1,\sqrt{s}}\,(\xmax)
\ M_{1,\sqrt{s}}\,(\xmin)
\, - {\pi \over 2} \, {x \, x_0 \, e^{-(x+x_0)/2}
\over 4 s - 1} \ .
\end{split}
\label{eq51}
\end{equation}
The right-hand side converges for all complex values of $s$ with
the exception of the point $s=1/4$ [which must be treated using
(\ref{eq46})] and the negative real semiaxis. The point $s=0$ is
convergent.

\eject

\vskip0.2truein
\centerline{\bf C. $\boldsymbol{\kappa = 2}$}
\vskip0.2truein

In this case there are two simple poles, located at $s_0 = 9/4$
and $s_1 = 1/4$. Utilizing the identity
\begin{equation}
\Gamma\left(-{3 \over 2} - i u \right) \,
\Gamma\left(-{3 \over 2} + i u \right) = {16 \pi \over \cosh(\pi u)
\, (9 + 4 u^2) \, (1 + 4 u^2)}
\ ,
\label{eq52}
\end{equation}
along with (\ref{eq48}), we can simplify (\ref{eq30}) to obtain
\begin{equation}
\begin{split}
\int_0^\infty
{u \, \sinh(\pi u) \over (9 + 4 u^2) \, (1 + 4 u^2) \, (s + u^2)}
\, W_{2 , \, i u}(x)
\, W_{2 , \, i u}(x_0) \, du \hskip2.0truein \\
= {\pi \over 32} \, {\Gamma(\sqrt{s}-3/2) \over
\Gamma(1 + 2 \sqrt{s})} \ W_{2,\sqrt{s}}\,(\xmax)
\ M_{2,\sqrt{s}}\,(\xmin) \hskip1.0truein \\
- \, {\pi \over 8} \, e^{-(x+x_0)/2} \, \sum_{n=0}^1
\, {(3-2\,n) \, n! \over \Gamma(4-n)}
\, {(x \, x_0)^{2-n} \over 4 s - (3-2\,n)^2}
\, P^{(3-2n)}_n(x) \ P^{(3-2n)}_n(x_0)
\ .
\end{split}
\label{eq53}
\end{equation}
Evaluation of the Laguerre polynomials yields
\begin{equation}
\begin{split}
\int_0^\infty
{u \, \sinh(\pi u) \over (9 + 4 u^2) \, (1 + 4 u^2) \, (s + u^2)}
\, W_{2 , \, i u}(x)
\, W_{2 , \, i u}(x_0) \, du \hskip2.0truein \\
= {\pi \over 32} \, {\Gamma(\sqrt{s}-3/2) \over
\Gamma(1 + 2 \sqrt{s})} \ W_{2,\sqrt{s}}\,(\xmax)
\ M_{2,\sqrt{s}}\,(\xmin) \hskip1.0truein \\
- {\pi \over 16} \ x \, x_0 \, e^{-(x+x_0)/2}
\, \left[{x \, x_0 \over 4 s - 9} + {(2-x) \, (2-x_0) \over
4 s - 1} \right]
\ ,
\end{split}
\label{eq54}
\end{equation}
which is convergent for all complex values of $s$, excluding the
negative real semiaxis and the points $s=1/4$, $s=9/4$. These two points
must be treated using (\ref{eq46}). Note that the point $s=0$ is
convergent in this case. Similar results can be obtained for any
positive or negative integer value of $\kappa$. The integral formula
given by (\ref{eq54}) is of particular significance in treating the
scattering of radiation in an ionized plasma with a constant
temperature, as discussed in Sec.~VII.

\section*{\bf VII. APPLICATION TO THERMAL COMPTONIZATION}

One of the most important physical applications of the results developed
in this paper involves the repeated Compton scattering of photons by a
hot Maxwellian distribution of electrons with temperature $T_e$ and
number density $n_e$ in an ionized plasma. This process, referred to as
``thermal Comptonization,'' is the primary mechanism responsible for the
production of the radiation spectra observed from celestial X-ray
sources such as active galaxies, black holes, and neutron stars.$^2$
When the electron temperature $T_e$ is constant, the Green's function,
$\green$, describing the temporal evolution of an initially
monoenergetic radiation distribution satisfies the Kompaneets partial
differential equation$^1$
\begin{equation}
{\partial \green \over \partial y} = {1 \over x^2} {\partial \over
\partial x} \left[x^4 \left(\green + {\partial \green \over \partial x}
\right)\right] \ ,
\label{eq55}
\end{equation}
where the dimensionless photon energy and the dimensionless time are
denoted by
\begin{equation}
x(\epsilon) \equiv {\epsilon \over k T_e} \ , \ \ \ \ \ \ 
y(t) \equiv \, n_e \sig c \, {k T_e \over m_e c^2} \, (t-t_0) \ ,
\label{eq56}
\end{equation}
respectively, and the quantities $\epsilon$, $t_0$, $t$, $\sig$, $m_e$,
$c$, and $k$ represent the photon energy, the initial time, the current
time, the Thomson cross section, the electron mass, the speed of light,
and Boltzmann's constant, respectively. The terms proportional to
$\green$ and $\partial \green / \partial x$ inside the parentheses on
the right-hand side of (\ref{eq55}) express in turn the effects of
electron recoil and stochastic (second-order Fermi) photon energization.
At the initial time $t=t_0$, the radiation distribution is monoenergetic,
and the Green's function satisfies the initial condition
\begin{equation}
\green(x,x_0,y) \Big|_{y=0} = x_0^{-2} \, \delta(x-x_0)
\label{eq57} \ ,
\end{equation}
where the dimensionless initial energy is given by
\begin{equation}
x_0 \equiv {\epsilon_0 \over k T_e} \ .
\label{eq58}
\end{equation}

By operating on (\ref{eq55}) with $\int_0^\infty x^2 \, dx$,
we can establish that $\green$ has the convenient normalization
\begin{equation}
\int_0^\infty x^2 \, \green(x,x_0,y) \, dx
= {\rm constant} = 1 \ ,
\label{eq59}
\end{equation}
where the final result follows from the initial condition
[Eq.~(\ref{eq57})]. Note that this normalization is maintained for all
values of $y$, which reflects the fact that Compton scattering conserves
photons. It can be shown based on (\ref{eq55}) that the Laplace
transform of the Green's function,
\begin{equation}
F(x,x_0,s) \equiv \int_0^\infty e^{-s y} \, \green(x,x_0,y) \, dy \ ,
\label{eq60}
\end{equation}
is given by$^3$
\begin{equation}
F(x,x_0,s) =
x_0^{-2} \, x^{-2} \, e^{(x_0-x)/2} \,
{\Gamma(\mu-3/2) \over \Gamma(1+2\mu)}
\ M_{2, \, \mu}(\xmin) \ W_{2, \, \mu}(\xmax) \ ,
\label{eq61}
\end{equation}
where the quantity $\mu$ is a function of the transform variable $s$,
defined by
\begin{equation}
\mu(s) \equiv \left(s + {9 \over 4}\right)^{1/2} \ ,
\label{eq62}
\end{equation}
and
\begin{equation}
\xmin \equiv \min(x,x_0) \ , \ \ \ \ \ \ 
\xmax \equiv \max(x,x_0) \ .
\label{eq63}
\end{equation}
The solution for the Green's function is obtained by performing
the inverse Laplace transformation using the Mellin integral,
\begin{equation}
\green(x,x_0,y) = {1 \over 2 \pi i} \, \int_{\gamma-i \infty}
^{\gamma+ i \infty} e^{sy} \, F(x,x_0,s) \, ds \ ,
\label{eq64}
\end{equation}
where the real constant $\gamma$ is chosen so that the line ${\mathcal
Re} \, s = \gamma$ lies to the right of the singularities in the
integrand. By transforming the variable of integration from $s$ to
\begin{equation}
s' \equiv s + {9 \over 4} \ ,
\label{eq65}
\end{equation}
we can obtain the equivalent expression
\begin{equation}
\green(x,x_0,y) = {e^{-9y/4} \over 2 \pi i} \, \int_{\gamma-i \infty}
^{\gamma+ i \infty} e^{s'y} \, \tilde F(x,x_0,s') \, ds' \ ,
\label{eq66}
\end{equation}
where
\begin{equation}
\tilde F(x,x_0,s') \equiv
x_0^{-2} \, x^{-2} \, e^{(x_0-x)/2} \,
{\Gamma(\sqrt{s'}-3/2) \over \Gamma(1+2\sqrt{s'})}
\ M_{2, \, \sqrt{s'}}\,(\xmin) \ W_{2, \, \sqrt{s'}}\,(\xmax) \ .
\label{eq67}
\end{equation}

The exact solution for the Green's function $\green(x,x_0,y)$ can
be obtained by taking the inverse Laplace transformation of (\ref{eq54}),
which yields
\begin{equation}
\begin{split}
{1 \over 2 \, \pi \, i} \, \int_{\gamma-i \infty}^{\gamma+ i \infty}
e^{sy} \, {\Gamma(\sqrt{s}-3/2) \over
\Gamma(1 + 2 \sqrt{s})} \ W_{2,\sqrt{s}}\,(\xmax)
\ M_{2,\sqrt{s}}\,(\xmin) \, ds \hskip2.0truein \\
= {32 \over \pi} \int_0^\infty
{u \, \sinh(\pi u) \over (9 + 4 u^2) \, (1 + 4 u^2)}
\, W_{2 , \, i u}(x)
\, W_{2 , \, i u}(x_0)
\ {1 \over 2 \, \pi \, i} \, \int_{\gamma-i \infty}^{\gamma+ i \infty}
{e^{s y} \over s + u^2} \ \, ds \ du \hskip0.50truein \\
+ \ {x x_0 \over 2} \ e^{-(x+x_0)/2} \ {1 \over 2 \, \pi \, i}
\, \int_{\gamma-i \infty}^{\gamma+ i \infty}
e^{s y} \left[{x x_0 \over s-9/4} + {(2-x)(2-x_0) \over s-1/4}
\right] \, ds \ ,
\end{split}
\label{compton1}
\end{equation}
where we have interchanged the order of integration in the double integral.
The inverse Laplace transformations on the right-hand side of (\ref{compton1})
are elementary in nature and can be evaluated using the formula
\begin{equation}
{1 \over 2 \pi i} \, \int_{\gamma-i \infty}^{\gamma+ i \infty}
{e^{sy} \over s+k} \, ds = e^{-k y} \ .
\label{compton2}
\end{equation}
By utilizing this result in (\ref{compton1}), we obtain
\begin{equation}
\begin{split}
{1 \over 2 \, \pi \, i} \, \int_{\gamma-i \infty}^{\gamma+ i \infty}
e^{sy} \, {\Gamma(\sqrt{s}-3/2) \over
\Gamma(1 + 2 \sqrt{s})} \ W_{2,\sqrt{s}}\,(\xmax)
\ M_{2,\sqrt{s}}\,(\xmin) \, ds \hskip1.5truein \\
= {32 \over \pi} \int_0^\infty
e^{-u^2 y} \, {u \, \sinh(\pi u) \over (9 + 4 u^2) \, (1 + 4 u^2)}
\, W_{2 , \, i u}(x)
\, W_{2 , \, i u}(x_0) \, du \hskip0.75truein \\
+ \ {x x_0 \over 2} \ e^{-(x+x_0)/2} \left[x x_0 \, e^{9y/4}
+(2-x)(2-x_0) \, e^{y/4} \right] \ .
\end{split}
\label{compton3}
\end{equation}
We can now combine (\ref{eq66}), (\ref{eq67}), and (\ref{compton3}) to
show that the exact solution for the time-dependent Green's function is
given by$^3$
\begin{equation}
\begin{split}
\green(x,x_0,y) =
{32 \over \pi} \ e^{-9y/4} x_0^{-2} x^{-2} e^{(x_0-x)/2}
\int_0^\infty e^{-u^2 y} \, {u \, \sinh(\pi u) \over
(1 + 4 u^2)(9 + 4 u^2)} \phantom{SPAAAAAACE} \\
\times \ W_{2, \, i u}(x_0) \,
W_{2, \, i u}(x) \, du
\ + \ {e^{-x} \over 2}
\ + \ {e^{-x-2y} \over 2} \ {(2 - x) \, (2 - x_0) \over
\ x_0 \, x} \ .
\end{split}
\label{eq68}
\end{equation}
Since the fundamental partial differential equation~(\ref{eq55}) is
linear, the particular solution for the radiation distribution
corresponding to an {\it arbitrary} initial spectrum can be found via
convolution using the Green's function. The result given by (\ref{eq68})
is therefore of central importance in the field of theoretical X-ray
astronomy.

\section*{\bf VIII. CONCLUSION}

In this paper we have developed several new formulas for the evaluation
of a family of integrals containing the product of two Whittaker
$W_{\kappa,\,\mu}(x)$-functions, when the integration occurs with
respect to the second index $\mu$, and that index is imaginary. The
fundamental integral we have focused on in this paper is
\begin{equation}
I(s) \equiv \int_0^\infty
{u \, \sinh(2 \pi u) \, \Gamma(1/2-\kappa-i u)
\, \Gamma(1/2-\kappa+i u) \over s + u^2}
\, W_{\kappa , \, i u}(x)
\, W_{\kappa , \, i u}(x_0) \, du \ .
\label{eq69}
\end{equation}
This is related to the Whittaker function index transformation discussed
in Refs.~12 and 13. An expression of particular interest is the
quadratic normalization integral given by (\ref{eq31}). The results
presented in (\ref{eq30}) and (\ref{eq46}) for $I(s)$ allow the exact
evaluation of all of the convergent cases of this integral without the
need to resort to numerical integration. We also point out that by
utilizing equations~(\ref{eq2}), one can easily obtain a set of
analogous integration formulas applicable to the Kummer functions
$\Phi(a,b,z)$ and $\Psi(a,b,z)$. While integrals of this precise type
have not been considered before, it is worth noting that $I(s)$ is a
member of a wider group of integrals containing the product of two
Whittaker $W$-functions. In general, the other integrals in this group
involve integration with respect to one of the other parameters, rather
than the second index as we have considered here. We briefly review a
few of these related integrals below.

Several formulas are available in the previous literature for evaluating
the integral of the product of two Whittaker $W_{\kappa,\,\mu}(x)$-functions
with respect to the primary argument $x$. For example, based upon
equation~(9.12) from Buchholz$^{10}$ or equation~(20.3.40) from
Erd\'elyi et al.$^{11}$ or equation~(7.611.3) from Gradshteyn and
Ryzhik,$^8$ we have
\begin{equation}
\begin{split}
\int_0^\infty W_{\kappa,\,\mu}(x) \, W_{\sigma,\,\mu}(x) \,
{dx \over x} = {1 \over \kappa-\sigma} \ {\pi \over \sin(2 \pi \mu)}
\ \bigg[{1 \over \Gamma(1/2 - \kappa + \mu) \Gamma(1/2 - \sigma - \mu)} \\
- \ {1 \over \Gamma(1/2 - \kappa - \mu) \Gamma(1/2 - \sigma + \mu)}
\bigg] \ ,
\hskip0.75truein
\end{split}
\label{eq70}
\end{equation}
which is valid provided $\abs{{\mathcal Re} \, \mu} < 1/2$. We note that
the formulas in Refs.~8 and 11 are missing a factor of $\pi$, and the
formula in Ref.~10 contains two incorrect signs. Another closely related
example is given by equation~(7.611.6) from Gradshteyn and Ryzhik$^8$ or
equation~(20.3.41) from Ref.~11,
\begin{equation}
\int_0^\infty x^{\sigma-1} \, W_{\kappa,\,\mu}(x) \, W_{-\kappa,\,\mu}(x)
\, dx = {\Gamma(\sigma+1) \, \Gamma(\sigma/2+1/2+\mu) \,
\Gamma(\sigma/2+1/2-\mu) \over 2 \ \Gamma(\sigma/2+1+\kappa)
\, \Gamma(\sigma/2+1-\kappa)}
\ ,
\label{eq71}
\end{equation}
which is valid provided ${\mathcal Re} \, \sigma > 2\,\abs{{\mathcal Re}
\, \mu}-1$.

A few formulas that treat the integration of a product of two Whittaker
$W_{\kappa,\,\mu}(x)$-functions with respect to the first index $\kappa$
have also been known for some time. The most general expression is
equation~(15.10b) from Buchholz,$^{10}$ which can be written as
\begin{equation}
\begin{split}
\int_0^\infty \Gamma(k-iu) \, \Gamma(k+iu) \,
W_{iu,\,k-1/2}(x) \, W_{-iu,\,k-1/2}(x_0) \, du
\hskip1.0truein \\
= \sqrt{\pi} \ \Gamma(2k) \, (x\,x_0)^k \, (x+x_0)^{-2k+1/2}
\, K_{2k-1/2}\left({x+x_0 \over 2}\right) \ ,
\end{split}
\label{eq72}
\end{equation}
where $K_{2k-1/2}(z)$ denotes the modified Bessel function. When
$k=1/2$, this formula reduces to equation~(7.691) from Ref.~8, which
states that
\begin{equation}
\int_0^\infty
{\rm sech}(\pi u)
\, W_{i u,\, 0}(x)
\, W_{-i u,\, 0}(x_0) \, du
= {\sqrt{x \, x_0} \over x + x_0} \ e^{-(x+x_0)/2} \ .
\label{eq73}
\end{equation}
The new results for $I(s)$ obtained in this paper [Eqs.~(\ref{eq30}) and
(\ref{eq46})] are in some sense complementary to these previously known
formulas. We emphasize that the expressions developed here are of
significance in a variety of applications, including the problem of the
Comptonization of radiation in an isothermal plasma, discussed in
Section~VII. Our general approach may also allow the determination of the
Green's function solution for the one-dimensional Schr\"odinger equation
with the Morse potential.$^{14}$ We plan to pursue this question in
future work.

The author would like to gratefully acknowledge the insightful comments
provided by the anonymous referee, which led to simplifications in the
main derivation and also helped to broaden the applicability of the results.

\section*{REFERENCES}

\smallskip\noindent
$^{1}$A. S. Kompaneets, ``The establishment of thermal
equilibrium between quanta and electrons,'' Sov. Phys. JETP {\bf 4},
730--737 (1957).

\smallskip\noindent
$^{2}$R. A. Sunyaev and L. G. Titarchuk, ``Comptonization
of X-rays in plasma clouds. Typical radiation spectra,'' Astron.
Astrophys. {\bf 86}, 121--138 (1980).

\smallskip\noindent
$^{3}$P. A. Becker, ``Exact solution for the Green's function
describing time-dependent thermal Comptonization,'' Monthly Notices
of the Royal Astron. Soc. {\bf 343}, 215--240 (2003).

\smallskip\noindent
$^{4}$D. Xianxi, J. Dai, and J. Dai, ``Orthogonality criteria
for singular states and the nonexistence of stationary states with even
parity for the one-dimensional hydrogen atom,'' Phys. Rev. A {\bf 55},
2617--2624 (1997).

\smallskip\noindent
$^{5}$V. V. Dodonov, I. A. Malkin, and V. I. Man'ko,
``The Green function of the stationary Schr\"odinger equation for
a particle in a uniform magnetic field,'' Phys. Letters {\bf 51},
133--134 (1975).

\smallskip\noindent
$^{6}$S. M. Blinder, ``Nonrelativistic Coulomb Green's
function in parabolic coordinates,'' J. Math. Phys. {\bf 22},
306--311 (1981).

\smallskip\noindent
$^{7}$V. Linetsky, ``Spectral expansions for Asian (average
price) options,'' Operations Research, in press (2003).

\smallskip\noindent
$^{8}$I. S. Gradshteyn and I. M. Ryzhik, {\it Table of Integrals,
Series, and Products} (Academic Press, London, 1980).

\smallskip\noindent
$^{9}$M. Abramowitz and I. A. Stegun, {\it Handbook of Mathematical
Functions} (Dover, New York, 1970).

\smallskip\noindent
$^{10}$H. Buchholz, {\it The Confluent Hypergeometric Function} (Springer,
New York, 1969).

\smallskip\noindent
$^{11}$A. Erd\'elyi, W. Magnus, F. Oberhettinger, and F. G. Tricomi,
{\it Tables of Integral Transforms} (McGraw-Hill,
New York, 1954), Vol. II.

\smallskip\noindent
$^{12}$Index Transforms: H. M. Srivastava, Yu. V. Vasil'ev, and S. B.
Yakubovich, ``A class of index transforms with Whittaker's function
as the kernel,'' Quart. J. Math. Oxford Ser. {\bf 49}, 375--394 (1998).

\smallskip\noindent
$^{13}$S. B. Yakubovich, ``Index transforms associated with products of
Whittaker's functions,'' J. Comp. Appl. Math. {\bf 148}, 419--427 (2002).

\smallskip\noindent
$^{14}$P. M. Morse, ``Diatomic molecules according to the wave
mechanics. II. Vibrational levels,'' Phys. Rev. {\bf 34},
57--64 (1929).

\end{document}